# Structure and Dynamics of the Coma Cluster

Matthew Colless[1]

Mount Stromlo and Siding Spring Observatories, The Australian National University,
Weston Creek, ACT 2611, Australia

and

Andrew M. Dunn

Institute of Astronomy, Madingley Road, Cambridge CB3 0HA, U.K.

## ABSTRACT

We examine the structure and dynamics of the galaxies in the Coma cluster using a catalog of 552 redshifts including 243 new measurements obtained with the Hydra multifiber spectrograph at KPNO. The velocity distribution for the 465 cluster members is clearly non-Gaussian. A new test for localised variations in the velocity distribution shows highly significant structure associated with the group of galaxies around NGC 4839, 40′ SW of the cluster core. We apply a mixture-modelling algorithm to the galaxy sample and obtain a robust partition into two subclusters: the main Coma cluster centered on NGC 4874, with $\overline{cz}{=}6853\,\mathrm{km\,s^{-1}}$ and $\sigma_{cz}{=}1082\,\mathrm{km\,s^{-1}}$, and a group centered on NGC 4839, with $\overline{cz}{=}7339\,\mathrm{km\,s^{-1}}$ and $\sigma_{cz}{=}329\,\mathrm{km\,s^{-1}}$.

We use this partition to examine the system's dynamics. We find that the velocity dispersion of the late-type galaxies in the main cluster is very close to $\sqrt{2}$ times that of the early-type galaxies, suggesting that the late types are freely-falling into a largely virialised cluster core dominated by early types. We obtain a virial mass for the main cluster of $0.9{\times}10^{15}\,h^{-1}\,M_\odot$, in close agreement with the estimates derived from recent X-ray data. The mass of the NGC 4839 group is found to be $0.6{\times}10^{14}\,h^{-1}\,M_\odot$, or about 5–10% the mass of the main cluster, in accord with their relative richnesses and X-ray luminosities. Assuming the main cluster and the NGC 4839 group follow a linear two-body orbit, the favored solution has the two clusters lying at 74 degrees to the line of sight at a true separation of $0.8\,h^{-1}\,\mathrm{Mpc}$ and moving together at $1700\,\mathrm{km\,s^{-1}}$.

Closer examination of the cluster core reveals an ongoing merger between two subclusters centered in projection on the dominant galaxies NGC 4874 and NGC 4889 but offset in velocity by $300\,\mathrm{km\,s^{-1}}$ and $1100\,\mathrm{km\,s^{-1}}$ respectively. Combining these results with X-ray and radio observations, and an interpretation of the presence or lack of an extended halo around the dominant galaxies NGC 4874, NGC 4889 and NGC 4839, we develop a merger history for the Coma cluster. We suggest that the

---

cD NGC 4839 is at the center of the group in which it formed, and that this group is just beginning to penetrate the Coma cluster, into which it is falling from the direction of Abell 1367 along the Great Wall. We argue that the radio halo of Coma and the disturbed X-ray emission in its core are the result of the ongoing merger between the main cluster and a group dominated by NGC 4889. This group has now been partially disrupted, ejecting NGC 4889. The position and apparent halo of NGC 4874 indicate that it was the original dominant galaxy of the main Coma cluster, though it may now have been dislodged from the bottom of the potential well.

## 1. INTRODUCTION

It is no longer possible to use Coma as the exemplar of a rich, regular and relaxed galaxy cluster. Studies of the projected distributions of both the galaxies and the X-ray gas show convincing evidence of statistically-significant substructures on both large and small scales, although evidence for substructure in the galaxy velocity distribution has been slight at best. In this paper we use new measurements of galaxy redshifts in Coma to demonstrate that the main structures visible in the projected maps are also identifiable in velocity space. We also take some initial steps towards understanding the dynamical significance of the observed substructure, and propose an outline of the merger history of the Coma cluster.

The most detailed and convincing evidence for substructure in Coma comes from the X-ray images obtained with ROSAT. Briel et al. (1992) construct a map of Coma using the ROSAT all-sky survey data which shows that the contours of the X-ray emission in the cluster core are elongated along the line joining the two central dominant galaxies, NGC 4874 and NGC 4889. The map also shows a secondary peak, accounting for 6% of the total cluster emission, around the group of galaxies centered on the cD galaxy NGC 4839, 40′ to the SW of the cluster core. A much deeper ROSAT image published by White et al. (1993) confirms these structures and suggests lower-level clumps of emission associated with a few of the other bright cluster galaxies.

The main structures seen in these X-ray maps had already been detected in the projected distribution of galaxies. Fitchett & Webster (1987) demonstrate the existence of double structure in the cluster core, with two clumps of galaxies centered on NGC 4874 and NGC 4889, while Mellier et al. (1988) use the color-magnitude relation for early-type galaxies to select probable cluster members and show that the large-scale galaxy distribution has a second peak around NGC 4839. Merritt & Tremblay (1994) apply an adaptive kernel method to obtain an improved map of the galaxy distribution based on Mellier et al.'s sample. A good summary of the evidence for substructure in Coma prior to the ROSAT maps is given by Baier et al. (1990).

Although structure is clearly apparent in the projected galaxy distribution, it is not so apparent in the distribution of radial velocities. Fitchett & Webster (1987) find only weak evidence of velocity structure associated with the two clumps they identify in the cluster core. Merritt (1987) notes that the velocity distribution of galaxies *outside* the cluster core is significantly skew, but Dressler & Shectman's (1988) test for localised departures from a Gaussian velocity distribution does not reveal any significant substructures.

This paper takes up the issue of velocity structure in the Coma cluster, inspired and guided by the beautiful ROSAT images. The analysis is based on a catalog of galaxy positions, magnitudes, colors and redshifts compiled from various sources in the literature, to which we have added more accurate positions and new redshifts for 243 galaxies. We use this catalog to examine the cluster's line-of-sight velocity distribution and find highly significant departures from a single Gaussian. Applying a powerful new test for localised velocity substructure, we are able to show that the origin of these departures is a group of galaxies centered on NGC 4839 and corresponding closely to the secondary peak in the ROSAT maps. We partition the galaxy sample between the main cluster and this NGC 4839 group using a maximum-likelihood algorithm, and derive mean velocities, velocity dispersions and masses for both subclusters. We model the dynamics of this system with a simple two-body model and obtain a favored solution supported by existing optical, X-ray and radio data. Looking for structure on smaller scales, we show the existence of velocity substructure associated with the two dominant galaxies NGC 4874 and NGC 4889 in the cluster core. Finally we attempt to synthesise all the available information on the structure and dynamics into a coherent picture of the merger history of the Coma cluster.

## 2. DATA

### 2.1. The Coma Redshift Catalog

Godwin, Metcalfe & Peach (1983, hereafter GMP) list 6724 galaxies brighter than $b=21$ in a square region of side 2.63 degrees (3.0 $h^{-1}$ Mpc) centered on the Coma cluster ($\alpha = 12^h59^m42\overset{s}{.}8$, $\delta = +27°58'14''$, J2000). The GMP sample is complete to $b=20$, at which limit it includes 2510 galaxies. Our catalog is based on this complete subsample of the GMP data.

The $b$ passband of GMP is defined by the combination of a IIIa-J emulsion and a Wr2C filter, zero-pointed with photoelectric $B$ band observations of 96 of the brighter galaxies. The $b$ magnitudes are integrated within the $\mu_b=26.5$ mag arcsec$^{-2}$ isophote. The $r$ passband is defined by the combination of a 127-04 emulsion and an RG1 filter, with a zero-point giving $b-r \approx B-R$. The $r$ magnitudes are integrated within the $\mu_r=24.75$ mag arcsec$^{-2}$ isophote. GMP claim external errors in $b$, $r$ and $b-r$ of approximately 0.15 mag.

The GMP catalog gives positions with an rms uncertainty of approximately 2″. This is inadequate for positioning fibers, so the GMP catalog was matched with the positions obtained from a UK Schmidt Telescope plate (J10027) scanned with the APM measuring machine at the Royal Greenwich Observatory. The APM positions have an rms uncertainty of less than 1″. Of the 2510 galaxies in the GMP catalog 2348 (94%) could be satisfactorily matched to an APM galaxy. Of the remaining 162 galaxies, 48 could not be matched within 6″ (roughly the joint $3\sigma$ uncertainty) and so the GMP position was kept, while in another 114 cases the match was doubtful because of a large difference between the GMP and APM magnitudes or because APM classed the object as a star or junk. Most of the cases of no or doubtful matching between the GMP and APM objects were fainter than $b=19$. The final set of positions were transformed from B1950 to J2000.

A near-complete set of redshifts from the literature were then obtained for the galaxies in this sample by searching the NED[1] database (as at 1 October 1993). Further redshifts were added to this list from Kent & Gunn (1982), van Haarlem et al. (1993) and Caldwell et al. (1993). The total number of redshifts found in this way was 345.

## 2.2. New Redshift Measurements

In order to supplement the literature observations, particularly in the regions of interest indicated by the ROSAT observations, we used the Hydra fiber spectrograph on the KPNO 4m telescope to measure new redshifts. The 40′-diameter Hydra fields we observed are shown in Figure 1, together with the GMP galaxies brighter than $b=20$ and the X-ray contours from White et al. (1993). Within each field we observed the brightest galaxies with $b<20$ that had no previous redshift measurement, plus a control sample of galaxies with known redshift.

The observations were made on 8–10 March 1992. Although most of the first two nights were lost to snow, we obtained 1.5–2 hours of integration on each of four fields. Spectra were obtained over 3700–5700 Å with a 632 lines/mm grating (KPC-007A) and Tektronix $2048^2$ CCD (T2KB), giving a resolution of 1.0 Å/pixel and 4.2 Å FWHM. The data were reduced using the standard IRAF tasks `ccdproc` and `dohydra`. Redshifts were obtained using the cross-correlation task `xcsao` for absorption-line objects and for emission-line objects by Gaussian fits to the line profiles.

The new redshifts are listed in Table 1. For each galaxy we give the GMP catalog number, the J2000 position, the $b$ magnitude and $b-r$ color from GMP, the heliocentric redshift $cz$ and its uncertainty $\Delta cz$ in $\mathrm{km\,s^{-1}}$, the Tonry & Davis (1979) $R$ value for the cross-correlation (for absorption-line spectra), a quality code $Q$ for the redshift measurement (described below), and the number of emission lines $E$ (if present).

The quality code was assigned to each redshift on the basis of visual inspection of both the spectrum and the cross-correlation function: $Q=0$ if no redshift identification could be made (the 110 $Q=0$ objects are omitted from Table 1); $Q=1$ if a cross-correlation redshift was obtained but was not certain (57 objects); $Q=2$ if the cross-correlation redshift was certain (161 objects); and $Q=9$ if a redshift was measured from emission lines (25 objects). The rms estimated uncertainties for quality codes $Q=1,2,9$ are $\delta cz=65,30,52\,\mathrm{km\,s^{-1}}$. Emission-line redshifts are considered certain if based on more than one line, as is the case for 22 of the 25 emission-line redshifts. As Figure 2 shows, the distinction between $Q=1$ and $Q=2$ corresponds quite closely to $R<3$ and $R>3$. Figure 3 shows some examples of $Q=1, 2$ and 9 spectra, with identified features marked.

The comparison with pre-1992 literature redshifts is limited, since we specifically tried to exclude such objects. Comparing our new results with the heterogeneous collection of redshifts found in NED, we find 15 objects in common, with a mean velocity difference of $25\,\mathrm{km\,s^{-1}}$ and an rms scatter of $122\,\mathrm{km\,s^{-1}}$ (see Figure 4a). A better external estimate of our velocity errors can be obtained from the 45 galaxies we have in common with the high-quality velocity measurements obtained by Caldwell et al. (1993). In this comparison we find an approximately Gaussian distribution for $\Delta cz$ with a slight offset in the mean of $-22\,\mathrm{km\,s^{-1}}$ and an rms scatter of only $48\,\mathrm{km\,s^{-1}}$, consistent with the quoted rms uncertainties (see Figure 4b). However these comparisons only apply to the brighter galaxies in our sample—all the objects involved are brighter than $b=19$ and have quality codes of either $Q=2$ or $Q=9$. The comparisons confirm the accuracy of the redshifts we consider certain, but provide no check on the less certain $Q=1$ measurements. Such a check is provided by some redshift measurements kindly communicated in advance of publication by Biviano et al. (1995). Of 52 objects in common, five have discrepancies of more than $500\,\mathrm{km\,s^{-1}}$ (GMP numbers 3080, 3768, 4659, 4749 and 4825). The remaining 47 objects have a mean offset of $-34\,\mathrm{km\,s^{-1}}$ and an rms scatter of $125\,\mathrm{km\,s^{-1}}$, dominated by the estimated error of approximately $100\,\mathrm{km\,s^{-1}}$ given by Biviano et al. for their velocities (see Figure 4c). Of the five discrepant objects, four have reliable ($Q=2$ or 9) ratings in our catalog but low quality ratings in Biviano et al.'s catalog, so we presume the error to lie in their estimates. Encouragingly, only one of the 12 $Q=1$ objects in the comparison has a discrepant redshift. We therefore conclude that our new redshifts are generally reliable to within the quoted errors.

---

[1] NED is the NASA/IPAC extragalactic database, operated for NASA by the Jet Propulsion Laboratory at Caltech.

To summarise, we have measured a total of 243 redshifts in the GMP region about the Coma cluster. These new measurements bring the total number of redshifts in the region around Coma to 552, making it by far the best-observed cluster in the sky. The following sections employ this redshift catalog to examine the structure and dynamics of the cluster.

## 3. CLUSTER STRUCTURE

### 3.1. Velocity Distribution

The line-of-sight distribution of $cz$ for the galaxies in our redshift catalog is shown in Figure 5. If we determine cluster membership using pessimistic $3\sigma$-clipping (Colless & Hewett 1987), we find there are 465 galaxies in the cluster, which spans the range 4000–$10000\,\mathrm{km\,s^{-1}}$. The mean and standard deviation for these galaxies are $\overline{cz}=6917\pm47\,\mathrm{km\,s^{-1}}$ (i.e. $\overline{z}=0.02307\pm0.00016$) and $\sigma_{cz}=1038\pm60\,\mathrm{km\,s^{-1}}$. The line-of-sight velocity dispersion of the cluster is $\sigma_v=\sigma_{cz}/(1+z)=1015\pm59\,\mathrm{km\,s^{-1}}$. In calculating the dispersion and the $1\sigma$ errors for the mean and dispersion we allow for the fact that the standard deviation of a $3\sigma$-clipped Gaussian is 0.983 times the true standard deviation; we also assume a typical error in $cz$ of $50\,\mathrm{km\,s^{-1}}$ and calculate the errors on $\overline{cz}$ and $\sigma_v$ according to Danese et al. (1980).

A range of statistical tests confirm the eye's impression (see Figure 5) that the velocity distribution departs significantly from a Gaussian. A $\chi^2$-test on the data binned as in Figure 5 gives $\chi^2=52.8$ with 26 degrees of freedom and rejects a Gaussian at the 99.8% confidence level;

the Lilliefors test (the Kolmorogorov-Smirnov test accounting for the fact that the mean and dispersion are estimated from the data) rejects a Gaussian at the 99.6% level; the Shapiro-Wilk $W$-test rejects a Gaussian at the 99.7% level. These departures from Gaussian form cannot, however, be simply characterised in terms of the low-order moments of the distribution: neither the skewness nor the kurtosis is very significantly non-Gaussian.

A much better match to the observed velocity distribution can be achieved with a double Gaussian. A $\chi^2$ fit gives $\overline{cz}_1$=6539±146 km s$^{-1}$, $\sigma_1$=1096±74 km s$^{-1}$ and $\overline{cz}_2$=7477±89 km s$^{-1}$, $\sigma_2$=476±93 km s$^{-1}$ with $\chi^2$=20.9 for 21 degrees of freedom. The greatly improved fit with a double Gaussian is the first clear suggestion that we are detecting substructure in the line-of-sight velocity distribution.

### 3.2. Localised Velocity Structure

Given the non-Gaussian nature of the velocity distribution and the second peak in the X-ray map around NGC 4839, we clearly should search for spatially localised variations in the line-of-sight velocity distribution. One test that has been commonly used for this purpose is the $\Delta$-test developed by Dressler & Shectman (1988). For each galaxy in the cluster they use the sample of 11 nearest neighbours (including the galaxy itself) to compute a local mean velocity $\overline{v}_{local}$ and velocity dispersion $\sigma_{local}$. They next compute the deviation $\delta$ of these values from the mean velocity $\overline{v}$ and dispersion $\sigma$ for the cluster as a whole using

$$\delta^2 = (11/\sigma^2)[(\overline{v}_{local} - \overline{v})^2 + (\sigma_{local} - \sigma)^2]. \tag{1}$$

The test statistic $\Delta$ is then the sum of the $\delta$'s over the $N$ galaxies in the cluster. Dressler & Shectman choose not to add the $\delta$'s in quadrature in order not to over-emphasise the largest deviations. For a Gaussian velocity distribution without local variations the expectation value of the statistic is $\langle\Delta\rangle=N$. The expected distribution of the statistic in the absence of localised deviations can, however, be estimated for any given cluster without this assumption by computing $\Delta$ for a large number of simulated clusters in which the galaxies have the same positions but have had their velocities randomly re-assigned. Dressler & Shectman used this statistic to look for substructure in a set of 15 clusters including Coma. Their test gave no evidence for substructure in Coma, but they used only 100 galaxies, of which only 3 were in the region of NGC 4839.

It is worthwhile to consider the nature of Dressler & Shectman's test before proceeding further. In essence it breaks down into three parts: (i) a recipe for defining what is 'local'; (ii) the comparison of one 1D distribution with another, its hypothetical parent; and (iii) a recipe for combining all the local comparisons to recover a single statistic for the whole cluster. The first point worth noting is that the statistic $\delta$ they adopt for performing part (ii) is not what one would *a priori* choose to compare two 1D distributions, especially if they are not necessarily Gaussian. The usual choices for this task are the $\chi^2$ or K-S $D$ statistics, which are more general (they apply to distributions that cannot be characterised by just the first two moments), more powerful (they are statistically more efficient, able to discriminate different distributions on the basis of smaller samples), and more robust (less susceptible outliers and pathological cases). Secondly, the scale imposed by the choice of the 11 nearest neighbours may not be the scale on which substructure is most obvious—clearly it is of interest to examine such statistics over a range of scales by allowing the number of nearest neighbours to vary. Finally, it would be helpful to have a statistic that was readily interpreted.

With these considerations in mind, we have devised a new test, in the spirit of the Dressler-Shectman test, for detecting localised variations in the velocity distribution. For each galaxy we compare the velocity distribution of the $n$ nearest neighbours to the velocity distribution of the whole cluster via a standard K-S two-sample test (see, e.g., Press et al., 1986, pp472-5). We define our test statistic $\kappa_n$ to be

$$\kappa_n = \sum_{i=1}^{N} -\log(P_{KS}(D > D_{obs})), \qquad (2)$$

where the sum is over the $N$ galaxies in the cluster and $P_{KS}(D > D_{obs})$ is the probability of the K-S statistic $D$ being greater than the observed value $D_{obs}$, which can be straightforwardly computed (see Press et al.). $\kappa_n$ is thus just the (negative) log-likelihood that there is no localised deviation in the velocity distribution on the scale of groups of $n$ nearest-neighbours—the larger $\kappa_n$, the greater the likelihood that the local velocity distribution is different from the overall distribution. As with the Dressler-Shectman statistic, the significance of $\kappa_n$ can be straightforwardly estimated by Monto Carlo simulations in which the velocities of the cluster galaxies are shuffled randomly.

Table 2 gives the results of applying this test to our catalog of redshifts in Coma. For various group sizes $n$, it gives the probability that $\kappa_n$ is greater than the observed value $\kappa_n^{obs}$ and the number of simulations used to determine this probability. It is clear that the velocity distribution in the Coma cluster has highly significant local variations for subgroups of galaxies with a broad range of sizes. The *location* of this localised variation is shown using groups of size $n=10$ and 50 in the bubbleplots of Figures 6 and 7. In these plots the size of the bubble at the position of each galaxy is proportional to $-\log(P_{KS}(D > D_{obs}))$, so that larger bubbles indicate a larger difference between the velocity distribution of the $n$ nearest-neighbours and the overall cluster velocity distribution. There are two clusters of large bubbles indicating two subclusters with differing velocity distributions. One is centered on the main body of the cluster and the other is centered around NGC 4839.

Figure 8 shows the velocity distributions in different regions of the cluster suggested by the above result: (a) within 20′ of the cluster center; (b) outside 20′ of the center; (c) within 20′ of NGC 4839; and (d) further than 20′ from both the cluster center and NGC 4839. The velocities of the three dominant cluster galaxies are indicated. For each of these four regions Table 3 gives the confidence level at which the skewness, K-S and $W$ tests reject a Gaussian distribution, together with the probability that each sample's velocity distribution is consistent with that of the sample within 20′ of the cluster center under a two-sample K-S test. As the figure shows, and the table confirms quantitatively, the deviation of the overall velocity distribution from a Gaussian can be almost entirely explained by the presence of a subcluster centered on NGC 4839.

A skewness in the velocity distribution outside the cluster core was first noted by Merritt (1987). Merritt & Gebhardt (1994) showed that this skewness was confined to distances between 16′ and 40′ from the cluster center, but did not identify it with the NGC 4839 subcluster. Once the region around NGC 4839 is excluded the skewness disappears, and the velocity distribution in the outer regions is found to be statistically consistent with that in the cluster core (Fig.8 and Table 3; cf. Fig.5 of Merritt & Gebhardt).

### 3.3. Characterisation of Subclusters

Having shown the existence of localised variations in the line-of-sight velocity distribution, and established that these are due to the galaxies around NGC 4839, we now attempt to characterise the membership and properties of the main cluster and the NGC 4839 subcluster.

A powerful tool for this task is the KMM mixture-modelling algorithm (McLachlan & Basford 1988), recently introduced to the study of galaxy clusters by Bird (1994). This algorithm attempts to find the maximum-likelihood partition of a dataset into a specified number $M$ of $N$-dimensional subclusters (see Ashman & Bird 1994). The $M$ subclusters are each assumed to have $N$-dimensional Gaussian distributions, with possibly different scalelengths in each dimension. The algorithm returns an estimate of the confidence level at which the $M$-cluster hypothesis is preferred to the null hypothesis of a single cluster, as well as the partition of the objects into $M$ Gaussian subclusters and the location and scale of each subcluster. The limitations of this algorithm are: (i) the lack of a decisive way to determine the optimum number of subclusters, and (ii) the necessary assumption that the subclusters are $N$-dimensional Gaussians. The former limitation is met by simply adopting the smallest number of subclusters that provides an adequate representation of the data. The latter is not too limiting for the study of clusters of galaxies, since each subcluster is supposed to be virialised and so have a Gaussian velocity distribution and a projected spatial distribution that is at least qualitatively similar to a Gaussian.

We here apply the KMM algorithm in three dimensions: the projected positions of the galaxies on the sky and the line-of-sight velocities. We use the 436 galaxies with $4000 < cz < 10000 \,\mathrm{km\,s^{-1}}$ lying within 70′ of the cluster center. We concentrate on a partition into just two groups, as this turns out to be optimum. The algorithm can start iterating towards the maximum-likelihood solution from either an initial partition of the sample into groups or from a set of initial parameters for each group. Table 4 lists the various initial parameters/partitions which we used and the corresponding final solutions the algorithm arrived at: $\langle x_1, y_1, v_1 \rangle$ and $\langle x_2, y_2, v_2 \rangle$ are the mean positions and velocities of the two subclusters, $\sigma(x_1, y_1, v_1)$ and $\sigma(x_2, y_2, v_2)$ are the standard deviations in these coordinates, and $f_1$ and $f_2$ are the fractions of the sample in the two subclusters. The table also gives the KMM algorithm's estimate of the overall rate for the correct allocation of objects to groups.

Given our previous results, we first look for a partition into two groups corresponding to the main body of Coma and the subcluster around NGC 4839. In case 1 we chose to specify initial positions and dispersions for the two subclusters (see Table 4), while in case 2 we specified a partition of the sample in which the NGC 4839 subcluster was taken to be all galaxies within $R=20'$ and $\Delta v=\pm 1200\,\mathrm{km\,s^{-1}}$ of NGC 4839. Although specified differently, the starting parameters for these two cases are quite similar and the final solutions reached were identical (i.e. an identical final partition into subclusters). The two-group solution was strongly favored compared to the one-group solution and the estimate of the correct allocation rate was 96%.

The critical question is how robust this final partition is under changes in the initial parameters. Cases 3–7 show that even initial guesses quite far from this solution converge to very similar results. In cases 3–5 the subclusters are defined as in case 2 but with the NGC 4839 subcluster having $R=15,25,30'$, $\Delta v=\pm 900,1500,2000\,\mathrm{km\,s^{-1}}$ and a fraction of galaxies $f_2=0.09,0.21,0.34$ respectively. Cases 3 and 4 converge to the exact same result as cases 1 and 2, while case 5 converges to a very similar result (the change is due to 6 galaxies being reallocated from the main cluster to the NGC 4839 subcluster). The initial parameters in cases 6 and 7 differ more radically from the best solution. In case 6 the positions and sizes are set to sensible values, but the mean velocities are reversed, the velocity dispersions for both subclusters are set to $1000\,\mathrm{km\,s^{-1}}$ and the fraction of galaxies in each is set to 0.5. Case 7 is the same as case 6 but this time both clusters are have their mean velocity set to $7000\,\mathrm{km\,s^{-1}}$—this set of initial parameters corresponds to using the projected distribution to separate the subclusters while making no assumptions about their velocity distributions. Even these two cases, with quite incorrect initial parameters, converge to the same solution as case 5. Only if we throw away the information on subclustering from the velocity distribution *and* the projected positions of the galaxies do we fail to achieve a similar solution. In case 8 we make the initial parameters for the two subclusters identical apart from a trivial symmetry-breaking difference in their positions. We then get a quite different result, with two oddly-placed and more elliptical subclusters and a significantly lower correct allocation rate.

The partition into two subclusters (the main Coma cluster and the NGC 4839 group) would therefore appear to be very robust. Figure 9 shows the final partition of the sample on the sky and in line-of-sight velocity corresponding to the solution for cases 1-4. It is instructive to compare the positions and velocities of the two subclusters, at (-0.9,-3.6,6853) and (-33.1,-29.0,7339), with those of the three dominant galaxies in the cluster: NGC 4874 at (-1.6,-0.7,7152), NGC 4889 at (5.6,0.4,6494), and NGC 4839 at (-30.7,-28.3,7352). The center of the main subcluster lies just 3.0′ from NGC 4874, while the center of the secondary subcluster lies just 2.5′ from NGC 4839. The mean velocity of the secondary subcluster is identical (within the measurement errors) to the velocity of NGC 4839. However NGC 4874 and NGC 4889 lie $+299\,\mathrm{km\,s^{-1}}$ and $-359\,\mathrm{km\,s^{-1}}$ respectively from the mean velocity of the main cluster (to which they clearly belong). Since the standard error in this mean is $111\,{\rm km\,s^{-1}}$, their peculiar velocities are statistically significant.

If we apply the same Gaussianity tests to these subclusters that we applied to the cluster as a whole, we find little evidence for non-Gaussian velocity distributions. The only statistic significant at >90% confidence is the $W$-test when applied to the main subcluster, which would appear to be picking up the deficit of galaxies with velocities around $6600\,{\rm km\,s^{-1}}$ that can be seen in Figure 9. In general, however, the velocity distribution in Coma is well-fitted by two Gaussians and there is no evidence for further significant subclustering.

The velocity parameters determined by this analysis give an accurate picture of the velocity structure in the cluster, since no observational biases were imposed on the measured redshifts. However the relative numbers of galaxies in the two subclusters and (to a lesser extent) the projected sizes are affected by the particular regions in which we chose to obtain redshifts (see Figure 1). We therefore do not obtain a reliable estimate of the relative richness of the subclusters from this analysis.

An alternative visualisation of the subclustering is provided by Figure 10, which shows the smoothed density of galaxies as a function of velocity and distance from the cluster center along the NE–SW diagonal (i.e. $(X+Y)/\sqrt{2}$, with NE positive). The latter coordinate was chosen so as to maximize the separation of the main cluster and the NGC 4839 subcluster. The smoothing was obtained with a 2D Gaussian kernel with a spatial dispersion of $8'$ and a velocity dispersion of $300\,{\rm km\,s^{-1}}$. These values were guided by physical scales in the cluster rather than by statistical considerations: $8'$ is approximately the cluster core radius, while $300\,{\rm km\,s^{-1}}$ is about the scale of the smallest velocity dispersions of groups of galaxies. The positions of the three dominant galaxies are indicated. The NGC 4839 group (with NGC 4839 close to its center) stands out clearly in the spatial coordinate but not in velocity (as Figure 9b implies). However this figure also shows interesting structure in the cluster core.

### 3.4. Core Structure

Figure 10 shows a correlation between position and velocity in the core of the main Coma cluster, with lower velocities to the SE and higher velocities to the NW. Given Fitchett & Webster's (1987) finding that there is double structure in the projected galaxy distribution in the core of Coma centered on NGC 4874 and NGC 4889, it is no surprise to see that these two dominant galaxies are projected in the spatial dimension onto the primary and seconday peaks, respectively, in the core galaxy distribution. Contrary to naive expectation, however, their velocities are *not* well-matched to the mean velocities of the peaks they are projected onto—NGC 4874 has a velocity about $350\,{\rm km\,s^{-1}}$ higher and NGC 4889 has a velocity about $1100\,{\rm km\,s^{-1}}$ lower. In fact the positions of the peaks and the dominant galaxies appear to be almost mirror-symmetric in velocity.

The significance of the correlation between position and velocity can be determined using rank correlation statistics. Taking the galaxies in a $30'\times30'$ central region with $6000\,\mathrm{km\,s^{-1}}<cz<8000\,\mathrm{km\,s^{-1}}$, we find that the Spearman $\rho$ and Kendall $\tau$ statistics imply correlations significant at the 98.5%, 79% and 99.3% confidence levels for $X$ vs $Y$, $X$ vs $cz$ and $Y$ vs $cz$; for $X+Y$ vs $X-Y$, $X+Y$ vs $cz$ and $X-Y$ vs $cz$ the correlations are significant at the 63%, 99.5% and 35% levels. Thus virtually all the correlation is in $X+Y$ vs $cz$ as shown in Figure 10. This trend is also illustrated in Figure 11, which shows four slices through the $X$–$Y$–$cz$ galaxy density distribution, smoothed using a 3D Gaussian filter with $(\sigma_X,\sigma_Y,\sigma_{cz})=(8',8',300\,\mathrm{km\,s^{-1}})$. In each slice the cluster core dominates, with the NGC 4839 group peaking strongly in the slice at $\overline{cz}=7250\,\mathrm{km\,s^{-1}}$. There is also a hint, at a much lower significance level, of a structure about $50'$ due east of the cluster center. However the feature in the figure that is relevant here is the shift in the centroid of the cluster core: at $6500\,\mathrm{km\,s^{-1}}$ the centroid is closer to NGC 4874 (which has $cz=7152\,\mathrm{km\,s^{-1}}$), at $7250\,\mathrm{km\,s^{-1}}$ it is midway between the two central dominant galaxies, and at $8000\,\mathrm{km\,s^{-1}}$ it is closer to NGC 4889 (which has $cz=6494\,\mathrm{km\,s^{-1}}$).

Applying the KMM algorithm to the galaxies in the cluster core is somewhat unsatisfactory. If no restriction is imposed on the velocity range the algorithm does not converge to a single well-defined final partition; if attention is limited to velocities in the range $6000\,\mathrm{km\,s^{-1}}<cz<8000\,\mathrm{km\,s^{-1}}$ then two subclusters with centers and mean velocities consistent with Figures 10 and 11 are recovered, but their sizes and velocity dispersions reflect mostly the imposed limits. Nonetheless, the two-group solution is strongly favored over the one-group solution, with one subcluster having NGC 4874 close to its projected center and a mean velocity of around $6800\,\mathrm{km\,s^{-1}}$, and the second having NGC 4889 close to its projected center and a mean velocity of around $7600\,\mathrm{km\,s^{-1}}$.

## 4. CLUSTER DYNAMICS

### 4.1. Field Galaxy Infall

The kinetic energy $K$ and gravitational potential energy $W$ of field galaxies falling freely onto a cluster are related by $|K/W|\approx1$, whereas galaxies in a virialised cluster core will have $|K/W|\approx1/2$. Thus the velocity dispersions of infalling and virialised galaxies are related by $\sigma_{infall}\approx\sqrt{2}\sigma_{virial}$. Huchra (1985) showed that the velocity dispersions of the spirals and elliptical galaxies in the Virgo cluster neatly accorded with this prediction, leading to the identification of the spirals as an infalling field population and the ellipticals as a virialised cluster population.

Lacking morphological classifications for the GMP sample we use the color-magnitude diagram to distinguish early- and late-type galaxies. Figure 12a shows the distribution of $b-r$ vs $b$ for the GMP sample, with its clearly defined color-magnitude relation for early-type galaxies. If we compare (Figure 12b) the surface density profiles of the early- and late-type galaxies brighter than $b=18$ (minimising the contribution of background objects), we see that (as expected) the early-type galaxies have a significantly steeper profile and dominate in the cluster core, where there is a large relative deficit of late-type galaxies.

The velocity distributions of the early- and late-type galaxies belonging (according to the KMM partition) to the main subcluster are shown in Figure 12c. The early types are consistent with the overall velocity distribution of the main subcluster (as they must be, given their dominance), but the late types have a velocity dispersion that is approximately $\sqrt{2}$ larger and statistically inconsistent with that of the early types—a two-sample K-S test shows that the distributions of the two types have only a 0.8% probability of being consistent. A similar analysis applied to the NGC 4839 subcluster finds that the velocity distributions of the early and late types there are consistent with each other, though the samples are much smaller. Although there is some suggestion in Figure 12c that some of the late types assigned to the main subcluster might actually belong to the NGC 4839 subcluster, the velocity distributions clearly support the hypothesis that the early types are a virialised population and the late types are still freely falling onto the cluster.

## 4.2. Cluster Mass

The best estimates of the mass of the Coma cluster now come from the X-ray data. Hughes (1989) combined the available X-ray and optical data to derive the mass inside $0.5\,h^{-1}$ Mpc and $2.5\,h^{-1}$ Mpc for a range of models. He found that the preferred models are those in which the mass follows the light, for which the mass inside $0.5\,h^{-1}$ Mpc is $M_{0.5}$=2.6–3.4$\times 10^{14}h^{-1}\,M_\odot$ and the mass inside $2.5\,h^{-1}$ Mpc is $M_{2.5}$=0.8–1.1$\times 10^{15}h^{-1}\,M_\odot$. Hughes also determined the minimum and maximum mass profiles consistent with all the data if the mass-to-light ratio of the data was not constrained to be a constant, obtaining allowed mass ranges of $M_{0.5}$=1.9–3.6$\times 10^{14}h^{-1}\,M_\odot$ and $M_{2.5}$=0.5–1.5$\times 10^{15}h^{-1}\,M_\odot$.

Briel et al. (1992) used their ROSAT survey image to measure the X-ray surface brightness out to 100′ from the cluster center. They find that the binding mass is more centrally concentrated than the X-ray gas, and obtain $M_{2.5}$=0.6–1.2$\times 10^{15}h^{-1}\,M_\odot$, with a fraction $0.11\pm 0.05\,h^{-3/2}$ of this mass in hot gas. They reduced the uncertainty in the cluster mass from the factor of 3 given by Hughes to a factor of 2 due to the fact that their X-ray surface brightness profile (derived excluding the region around NGC 4839) reaches twice as far out as any previous measurement and so provides a stronger constraint on the mass profile at large radii. Another mass estimate is given by Watt et al. (1992), who obtained an X-ray image of Coma with a telescope flown on Spacelab-2. They give $M_{0.5}$=2.1–2.6$\times 10^{14}h^{-1}\,M_\odot$ and a best-fitting total mass of $1.3\times 10^{15}h^{-1}\,M_\odot$. Both Briel et al.'s and Watt et al.'s mass estimates are consistent with Hughes' earlier values, and with the masses obtained from the optical galaxy data by Merritt (1987) and The & White (1986). More recent analyses of the galaxy data which use the line-of-sight velocity distribution to constrain the relative distributions of the dark and luminous matter (Merritt & Saha 1993) or take a non-parametric approach to reconstructing the density profile (Merritt & Gebhardt 1994) also yield similar results, while emphasising the large uncertainties in determinations of the mass distribution from the galaxy kinematics alone.

Although the X-ray mass determinations remain superior, we can use our expanded sample of galaxy redshifts and our partition of this sample into the main Coma cluster and the NGC 4839 group to obtain improved optical estimates for the total masses of these two components. The virial mass estimator gives $0.9 \times 10^{15} h^{-1} M_\odot$ for the main cluster and $0.6 \times 10^{14} h^{-1} M_\odot$ for the NGC 4839 group. Some estimate of the range of allowed masses (assuming mass follows light) is given by using the projected mass estimator (Bahcall & Tremaine 1981, Heisler et al. 1985) assuming isotropic and linear orbits, which probably under- and over-estimate the mass respectively. For these estimators we find total masses of $0.6$–$1.1 \times 10^{15} h^{-1} M_\odot$ for the main cluster and $0.5$–$1.0 \times 10^{14} h^{-1} M_\odot$ for the NGC 4839 group. These estimates for the mass of the main cluster are in good agreement with all the X-ray estimates. The mass of the NGC 4839 group is 5–10% of the mass of the main cluster, broadly consistent with the relative numbers of galaxies and the relative X-ray luminosities. Although Hughes (1989) finds that the preferred model is one in which mass follows light, it should be noted that a broader class of models in which the radial profiles of the mass-to-light ratio and the galaxies' orbital anisotropy vary together are also consistent with the data and yield a wider range of mass estimates.

### 4.3. Two-body Models

The simplest dynamical model for the Coma cluster/NGC 4839 group system is the two-body model of Peebles (1971), first applied to clusters by Beers et al. (1982) and Gregory & Thompson (1984). In this model the two clusters act as point masses following a linear orbit under their mutual gravity. The clusters are presumed to have started with zero separation and then to have moved apart before turning around and coming together again (the model assumes that the clusters are moving apart or together for the first time). Given the projected separation of the two clusters $R_p$, their relative radial velocity $V_r$, and their total mass $M$, the model yields the projection angle of the system $\alpha$, the true three-dimensional separation $R$ and relative velocity $V$, and the mass required to bind the system $M_{bind}$ (note that $\alpha$ is independent of $H_0$ while the radii and masses scale as $h^{-1}$). Because of the ambiguity inherent in seeing the system only in projection, the model usually leads to more than one orbital solution, corresponding to various values of $\alpha$ (the angle between the line joining the two clusters and the plane of the sky).

The two-body model neglects any angular momentum the system may have, ignores any matter outside the two subclusters, and does not allow for the distribution of matter within the individual clusters (which will become important when they are close to merger). Nonetheless it provides the best available description for double cluster systems given the limited dynamical constraints. It has previously been applied to a number of double clusters by Colless (1987) and Beers et al. (1991).

The input parameters for a two-body model of the Coma cluster and NGC 4839 group are provided by the KMM analysis (Table 4), which gives $R_p=0.81 \pm 0.06\,h^{-1}$ Mpc (for $q_0=0.5$) and $V_r=475 \pm 71\,\mathrm{km\,s^{-1}}$ [note that the physical line-of-sight velocity difference is $\Delta cz/(1+z)$]. These inputs lead to the set of solutions shown as the thick curve in Figure 13, which relates the projection angle to the total mass of the system. The upper and lower thin curves in the figure correspond to $V_r=404\,\mathrm{km\,s^{-1}}$, $R_p=0.75\,h^{-1}$ Mpc and $V_r=546\,\mathrm{km\,s^{-1}}$, $R_p=0.87\,h^{-1}$ Mpc respectively, showing the range of solutions allowed by the uncertainties in these quantities. To completely solve the model we require the total mass of the system. Since the main Coma cluster completely dominates the NGC 4839 group (see above), the appropriate mass to use is the mass of the main cluster inside the radius corresponding to the true separation $R$ of two clusters. Since $R$ depends on $\alpha$, we need to know the mass of the main cluster as a function of radius. For the purposes of this analysis it is sufficient for us to approximate the mass profile as a power law $\log M(<R) \propto \log R$ constrained to pass through the mass estimates at $0.5\,h^{-1}$ Mpc and $2.5\,h^{-1}$ Mpc as given in the previous section. The mass enclosed within the radius of the NGC 4839 group is shown in Figure 13 as a function of projection angle $\alpha$: the heavy shading corresponds to mass profiles in which mass follows light, while the unconstrained mass profiles are indicated by the lighter shading.

Acceptable solutions for the two-body model thus lie within the regions bounded by the thin curves and the light shading; the preferred solutions are where the thick curves intersect the heavy shading. These solutions are listed in Table 5. There are no unbound solutions—the NGC 4839 group is clearly bound to Coma—but there are three bound solutions, one outgoing (the system has not yet reached turnaround) and two incoming. The bound outgoing solution (BO) has the system aligned at $5^{+1}_{-1}$ degrees to the line of sight, with the NGC 4839 group lying $8.9^{+2.3}_{-1.0}\,h^{-1}$ Mpc directly behind the Coma cluster and moving away from it at $480^{+70}_{-470}\,\mathrm{km\,s^{-1}}$. The first bound incoming solution (BI1) has the system aligned at $10^{+8}_{-4}$ degrees to the line of sight, with the NGC 4839 subcluster closer to us than the main Coma cluster by $4.5^{+2.4}_{-1.7}\,h^{-1}$ Mpc and infalling with a velocity of $480^{+90}_{-80}\,\mathrm{km\,s^{-1}}$. The second bound incoming solution (BI2) has the system aligned at $74^{+5}_{-10}$ degrees to the line of sight, with the main cluster and the NGC 4839 subcluster separated by $0.8^{+0.1}_{-0.1}\,h^{-1}$ Mpc and moving together at a velocity of $1700^{+350}_{-500}\,\mathrm{km\,s^{-1}}$.

We can estimate the relative probability of these solutions if we assume that all dynamically-allowed orientations of such a system are equally likely, for then the probability of observing the system at projection angles in the range $\alpha_1$ to $\alpha_2$ is proportional to $\sin\alpha_2 - \sin\alpha_1$. Using the allowed ranges of $\alpha$ for the three solutions given in Table 5, we find that probabilities of the BO, BI1 and BI2 solutions are 1%, 15% and 84% respectively. Thus it is far more likely that the NGC 4839 system is incoming rather than outgoing. Of the two incoming solutions, BI2 is significantly favored over BI1. The preferred model for the system therefore has the two clusters lying at 74 degrees to the line of sight with a true separation of about $0.8\,h^{-1}$ Mpc and moving together at around $1700\,\mathrm{km\,s^{-1}}$.

## 5. DISCUSSION

### 5.1. Multiple Substructure and Merging in Coma

There is now strong evidence from X-ray, optical and radio observations for multiple substructure and ongoing merging in Coma.

1. The ROSAT images by Briel et al. (1992) and White et al. (1993) show that the core of the cluster has an elongated structure joining the two dominant galaxies NGC 4874 and NGC 4889. There is also a secondary concentration producing 6% of the total X-ray emission centered on NGC 4839, 40′ SW of the cluster core (see also Watt et al. 1992). The luminosity and the temperature of this subcluster are similar to those of poor clusters with central dominant galaxies (cf. Kriss et al. 1983). Evidence that the NGC 4839 group is infalling is provided by the way the X-ray emission trails outward from the cluster core behind NGC 4839. There is also some indication of structure associated with NGC 4911, 20′ SE of the core. Coma has no cooling flow (Edge et al. 1992), probably because it has been disrupted in a merger event (McGlynn & Fabian 1984). The lack of strong temperature and metallicity gradients in the X-ray gas in Coma could also be due to a major merger (Watt et al. 1992).

2. The projected galaxy distribution reflects the same structures as the gas, with a secondary peak around NGC 4839 (Mellier et al. 1988) and two clumps centered on NGC 4874 and NGC 4889 in the core (Fitchett & Webster 1987, Baier et al. 1990). Again there is some evidence for a substructure associated with NGC 4911 (Mellier et al. 1988).

3. The analysis of the galaxy velocities presented here clearly shows that the NGC 4839 group is a separate dynamical entity falling into the main cluster (§3.2 and §3.3). It has about 10% the mass of the main cluster (§4.2). A simple two-body model (§4.3) suggests that it is currently just beginning to penetrate the main cluster and will encounter the cluster core in less than 0.5 Gyr. There is also a significant correlation between the positions and velocities in the cluster core (§3.4), with two apparent subclusters separated by only 5.5′ on the sky but by $800\,\mathrm{km\,s^{-1}}$ along the line of sight. In projection these subclusters are associated with the two dominant galaxies NGC 4874 and NGC 4889, although they are offset in velocity by $350\,\mathrm{km\,s^{-1}}$ and $1100\,\mathrm{km\,s^{-1}}$ respectively.

4. The radio halo of the Coma cluster (Coma C; Venturi et al. 1990, Kim et al. 1990) is one of only a few known; all are in clusters without cooling flows (Burns et al. 1992). Such halos require a source of relativistic electrons, a large-scale magnetic field and in situ acceleration of the electrons. Tribble (1993) suggests that recent merging of subclusters may be necessary to create the conditions required for a persistent cluster-wide radio halo (see also Burns et al. 1994). The NGC 4839 group lies outside the radio halo of Coma, but NGC 4839 itself is a head-tail radio source with the tail pointing away from the main cluster (Venturi et al. 1990), implying that it is falling into the cluster through a fairly dense intracluster medium.

5. There is an excess of early-type galaxies showing evidence of recent star-formation in the region of NGC 4839, particularly between NGC 4839 and the cluster core (Caldwell et al. 1993). The star-formation in these objects might plausibly have been triggered by interaction with either the mean tidal field or the intracluster medium as they fell into the main cluster. However, as Caldwell et al. note, it is "difficult to maintain that the abnormal-spectrum galaxies are all gravitationally bound members of the SW substructure" since although the 17 abnormal galaxies in the SW region have an appropriate mean velocity of $7288\,\mathrm{km\,s^{-1}}$, their velocity dispersion is $1262\,\mathrm{km\,s^{-1}}$ (cf. the velocity dispersion of $329\,\mathrm{km\,s^{-1}}$ for the NGC 4839 group obtained here). And indeed only 3 of these galaxies are assigned to the NGC 4839 group by the KMM partition.

## 5.2. The Dominant Galaxies

The properties of the dominant galaxies NGC 4874, NGC 4889 and NGC 4839, especially their extended halos (or lack thereof), are particularly revealing about the substructure and merger history of Coma. Dominant galaxies are believed to form in poor clusters and groups, where the low velocity dispersion facilitates the mergers needed for them to grow (Malumuth 1992 and references therein). In this process an extended halo also forms out of stellar debris from the merger process and tidal stripping of other galaxies. This halo material, which distinguishes a cD from a D galaxy, orbits in the potential well of the entire group and is not bound to the galaxy itself. When the group merges with a rich cluster and is disrupted the halo is destroyed and the cD becomes a D galaxy. Tidal friction causes the very massive ($\sim 10^{13}\,h^{-1}\,M_\odot$) D galaxy to spiral to the cluster center, where it merges with any existing D or cD. Eventually it may re-acquire a halo from merger debris and again become a cD.

Schombert (1988) finds clear evidence for extended halos around both NGC 4874 and NGC 4839 and classifies them as cD galaxies. However there is no evidence for such a halo around NGC 4889, which is therefore classified as a D galaxy. Baier et al. (1990) note several pieces of evidence supporting the interpretation that NGC 4874 is at the bottom of the cluster potential well while NGC 4889 belongs to a recently accreted and disrupted group: (i) NGC 4874 lies at the peak of the X-ray emission and the center of the emission contours at larger scales; (ii) the number of galaxies close to NGC 4874 is 50% greater than around NGC 4889; (iii) NGC 4874 is a strong radio source coincident with the peak of the extended radio halo of the cluster—by contrast NGC 4889 is not distinguishable in the radio halo. Both galaxies have masses (within $80\,h^{-1}$ kpc) of approximately $1.4\times 10^{13}\,h^{-1}\,M_\odot$ (Vikhlinin et al. 1994).

With the new information from our analysis of the velocity structure in Coma, one can plausibly argue that, in terms of the evolutionary sequence for dominant galaxies outlined above, NGC 4839 is a cD at the center of the group in which it originally formed, while NGC 4889 is a D galaxy which lost any halo it may have had when it was ejected from its parent group (the core subcluster at $7600\,\mathrm{km\,s^{-1}}$) in the ongoing merger with the main Coma cluster. The case of NGC 4874 is less clear. It may be a genuine cD at the center of the main cluster, as argued above, though with a peculiar velocity of $300\,\mathrm{km\,s^{-1}}$ with respect to the peak of the galaxy distribution. Alternatively, it may be a D galaxy which is coincidentally projected onto the halo existing at the bottom of the cluster's potential well. In the latter case it may have been ejected from the cluster center in the same encounter that ejected NGC 4889 from the center of its own group.

Thus Coma provides a remarkable case in which various phases in the evolution of D/cD galaxies can be seen simultaneously. Furthermore the evolutionary stage of the dominant galaxies gives a strong indication of the merger stage of their parent group with the main cluster: the NGC 4839 group is just encountering Coma and is yet to be significantly disrupted, the NGC 4889 'group' has been partially disrupted but has not yet fully merged, while NGC 4874 is the original dominant galaxy of the main cluster, which may have been dislodged from its position at the bottom of the potential well by the encounter with NGC 4889. A similar picture has been proposed by White et al. (1993) to explain the structure in the X-ray gas.

### 5.3. Is Coma Pre- or Post-Prandial?

It has recently been suggested by Burns et al. (1994) that the Coma cluster has just had the NGC 4839 group for lunch: in other words, that the NGC 4839 group has already passed through the core of the main Coma cluster. This is counter to the picture developed above. Burns et al.'s main arguments (and the counter-arguments) are as follows.

1. The main cluster's X-ray surface brightness distribution, its isothermal temperature profile, its lack of a cooling flow, and its radio halo all argue strongly for a recent merger event. Burns et al. argue that the NGC 4839 group's passage through the cluster core was that event, however the arguments of the previous two sections would point to the ongoing merger involving NGC 4889 as a more likely culprit.

2. Burns et al. claim that the X-ray emission around NGC 4839 is unlike that expected for a bound group which has just begun to fall into the main cluster because of the relative flatness of the X-ray profile. They also suggest that the compact X-ray core about NGC 4839 is due to a cooling flow. However a cooling flow would not survive a merger—indeed Burns et al. argue that the *lack* of a cooling flow in the main cluster is due to the passage of the NGC 4839 group. The distorted outer X-ray emission contours could more plausibly be interpreted as the trailing plume produced by the group's passage through the intracluster medium of the main cluster (Briel et al. 1992, White et al. 1993).

3. Caldwell et al. (1993) obtained a velocity dispersion of $963\,\mathrm{km\,s^{-1}}$ for galaxies in the SW region of Coma, which Burns et al. correctly point out is inconsistent with a pre-merger bound group. However the analysis of this paper shows that there *is* a bound group about NGC 4839 with an appropriately small velocity dispersion; the large value obtained by Caldwell et al. is simply due to the mixture of galaxies from both the main cluster and the NGC 4839 group present in the SW region (see Figure 9). González-Casado et al. (1995) show using both analytic arguments and numerical simulations that accreted groups and small clusters are tidally disrupted in one cluster crossing (as is the group in Burns et al.'s simulation). Thus the existence of a bound group around NGC 4839 argues strongly against it having already passed through the cluster core.

4. Burns et al. note the existence of a filament of galaxies running NE from Coma towards the Zwicky cluster Zw 1319.9+3135, and suggest that the NGC 4839 group may have fallen into Coma from along this filament, based on a supposed radial velocity for the group that is $100\,\mathrm{km\,s^{-1}}$ smaller than that of the main cluster. However this filament is part of the Great Wall that extends through Coma to the SW towards the Abell cluster A1367 lying $21\,h^{-1}\,\mathrm{Mpc}$ away. From the results obtained here, the NGC 4839 group has a radial velocity $500\,\mathrm{km\,s^{-1}}$ *greater* than that of the main cluster. Using the position and redshift of A1367 to define the direction of the Great Wall, we find it runs away from Coma towards A1367 at a position angle on the sky of 36 degrees south of west, and at a projection angle $\alpha$ of 19 degrees. This is in good agreement with the observed position angle of the NGC 4839 group and the projection angle corresponding to the favored BI2 solution of our two-body model.

5. Burns et al. note that the distribution and velocities of the early-type galaxies with abnormal spectra found by Caldwell et al. (1993) in the SW region are more like the 'spray' of galaxies seen post-merger in their simulation than like the members of a bound group. The present analysis agrees in finding that few of these objects are likely to be members of the NGC 4839 group. However an alternative explanation for these galaxies is that, while not bound to the NGC 4839 group, they too are recent arrivals at the Coma cluster, onto which they have fallen (like NGC 4839) along the Great Wall from the SW. That such infall occurs is indicated by the broad velocity distribution seen for the bluer galaxies in Figure 12c. Moreover early-type *field* galaxies are more likely to have retained their gas reserves than those in the relatively dense environment of the NGC 4839 group. They are therefore more plausible candidates for a burst of star-formation. In this context it is clearly of interest to see if there are significant numbers of abnormal-spectrum galaxies at a similar distance from the cluster core but to the NE, arriving from the other side of the Great Wall (this region was not studied by Caldwell et al.).

## 6. CONCLUSIONS

Inspired by the recent ROSAT images of the Coma cluster to re-examine the structure of the galaxy distribution, we have measured redshifts for 243 galaxies using the Hydra multifiber spectrograph at KPNO. Adding these to literature redshifts, we have compiled a catalog of 552 redshifts for Coma based on the photometric sample of Godwin et al. (1983), almost doubling the number of cluster members with redshifts and providing much improved coverage of the regions showing structure in the ROSAT images.

The velocity distribution for the 465 cluster members is patently non-Gaussian. We introduce a new test for localised departures from a Gaussian velocity distribution which shows highly significant structure associated with the group of galaxies around NGC 4839, 40′ SW of the cluster core. We apply the KMM mixture-modelling algorithm to the galaxy sample and obtain a robust partition into two subclusters: the main Coma cluster centered on NGC 4874, with $\overline{cz}$=6853 km s$^{-1}$ and $\sigma_{cz}$=1082 km s$^{-1}$, and a group centered on NGC 4839, with $\overline{cz}$=7339 km s$^{-1}$ and $\sigma_{cz}$=329 km s$^{-1}$.

We use this partition to examine the system's dynamics. Using only the galaxies assigned to the main cluster, we compare the velocity distributions of the red, early-type galaxies and the blue, late-type galaxies and find that $\sigma_{late} \approx \sqrt{2}\sigma_{early}$, suggesting that the late types are freely-falling into a virialised main cluster dominated by early types. We obtain a virial mass for the main cluster of $0.9 \times 10^{15} h^{-1} M_\odot$, in excellent agreement with the best estimates derived from recent X-ray data. The mass of the NGC 4839 group is found to be $0.6 \times 10^{14} h^{-1} M_\odot$, or about 5–10% the mass of the main cluster, in accord with their relative richnesses and X-ray luminosities.

Assuming the main cluster and NGC 4839 group follow a linear two-body orbit, we obtain three possible solutions for the system's current state (the ambiguity results from only seeing the system in projection). All three solutions have the NGC 4839 group bound to the main cluster, with the most highly-preferred solution having the two clusters lying at 74 degrees to the line of sight at a true (three-dimensional) separation of $0.8\,h^{-1}$ Mpc and moving together at 1700 km s$^{-1}$.

A closer look at the cluster core shows a highly significant correlation between the galaxy positions and velocities. Plots of the 3D galaxy density and a KMM mixture model for the core suggest the presence of two subclusters: a true central peak close to the projected position of NGC 4874, but with $\overline{cz} \approx 6800$ km s$^{-1}$, and a secondary peak close to the projected position of NGC 4889, but with $\overline{cz} \approx 7600$ km s$^{-1}$.

Taking these results in conjunction with other optical, X-ray and radio observations, we conclude that there is now overwhelming evidence for multiple substructure and ongoing merging in the Coma cluster. We argue that the positions and velocities of the three dominant galaxies, and in particular their possession or lack of an extended halo (i.e. whether they are a cD or a D galaxy), reveal the current status and merger history of the cluster. In our picture, the cD NGC 4839 is at the center of the group in which it originally formed; this group is just falling onto Coma and is yet to be significantly disrupted (contrary to a recent suggestion by Burns et al. 1994). The disturbed X-ray emission and radio halo in the core of the cluster result from the ongoing merger of Coma with a group or cluster of which NGC 4889 was the dominant member; this group has already been partially disrupted (ejecting NGC 4889 and causing the loss of any halo it may have had) but is not yet fully assimilated. NGC 4874 is the true dominant galaxy of the main cluster—it appears to possess an extended halo and lies at the projected peak of the galaxy distribution and the X-ray and radio emission. However it does not lie at the mean velocity of the main cluster, so it may conceivably have been dislodged from the cluster center in an encounter with NGC 4889. In that case it may be a D galaxy projected onto the halo still residing at the bottom of the cluster potential well.

The future evolution of Coma is likely to be as turbulent as its past. NGC 4874 and NGC 4889 will eventually merge to form a single cD in the core of the cluster, only to be disturbed once more by the impact of the NGC 4839 group. The final dynamical relaxation of the cluster is indefinitely postponed.

We thank Simon White for suggesting follow-up of the ROSAT results and Gus Oemler for assistance with the observations. Christina Bird kindly provided us with the KMM code and helpful advice on how to use it. Agris Kalnajs made suggestions on the the use of density maps to probe for substructure which materially improved our analysis.

---